\documentclass{physeauth}
\usepackage{graphicx}
\usepackage{amsmath}
\usepackage{amssymb}

\begin{document}
\begin{frontmatter}
\title{Aspects of quantum cooling in electron and atom systems}
\author{Fernando Sols
\thanksref{thank1}}

\address{Departamento de F{\'\i}sica de Materiales, Universidad
  Complutense de Madrid, E-28040 Madrid, Spain.}

\thanks[thank1]{e-mail: f.sols@fis.ucm.es}

\begin{abstract}
Some features of nonadiabatic electron heat pumps are studied and connected to general questions of quantum cooling. Inelastic reflection is shown to contribute to heating if the external driving signal is time-symmetric. The quantum of cooling power, $\pi^2 k_B^2 T^2/6h$, is shown to be an upper limit to the cooling rate per transport channel in the presence of an arbitrary driving signal. The quantum limit to bulk atom cooling is also discussed. Within the electron tunneling limit, it is shown that electron cooling still occurs if the coherent ac source is replaced by a sufficiently hot thermal bath. A comparison with related refrigeration setups is presented.
\end{abstract}

\end{frontmatter}

\section{Introduction.}

The generation and flow of heat is a most important issue for
the increasingly miniaturized modern electronics \cite{cahi03,giaz06}.
The quantum of thermal
conductance, which is independent of the carrier
statistics \cite{rego99}, has been recently measured for phonons
\cite{schw00} and photons \cite{mesc06}. A practical and fundamental
issue is the identification of possible cooling mechanisms for
electron systems, a subject less developed than its atom
counterpart \cite{metc99}.
Heat pumping may be viewed as a particular instance of motion rectification \cite{reim02,marc09}.
Adiabatic electron \cite{hump02,peko06}
and molecular \cite{sega06} pumps may provide reversible heat
engines which would cool with minimum work expenditure. It has also
been proposed and shown that normal-superconductor interfaces can efficiently
cool the normal metal under appropriate conditions of electron
flow \cite{nahu94,clar05}. Within such a context, heat pumping might be enhanced
by extracting energy from a hot Ohmic resistor \cite{peko07}. More recently, refrigeration of a two-dimensional electron gas has been realized by using quantum dots to filter the energy of the current-carrying electrons \cite{pran09}.

\begin{figure}[tbp]
\begin{center}
\includegraphics{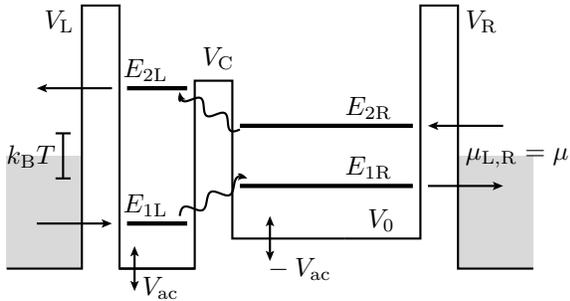} 
\end{center}
\caption{Asymmetric double-well heterostructure where the dominant transmission processes
contribute to cooling: in lead R \textit{hot} electrons
are replaced by \textit{cold} electrons, all within a range $\sim
k_{B}T$ around $\mu$. From Ref. \cite{rey07}.} \label{eq:model-processes}
\end{figure}

An alternative electron cooling mechanism has been proposed which would operate at zero electric current by exchanging hot for cold electrons at the interface with a warmer electrode \cite{rey07}. Such a pumping of heat would be driven nonadiabatically by an external ac source and the electron energy would be selected through a intermediate resonant structure.
The cooling concept is schematically depicted in Fig.~\ref{eq:model-processes}. In the present work, I expand on the content of Ref. \cite{rey07} by providing some mathematical proofs and by extending the discussion to include bulk atom cooling and some complementary questions of electron cooling such as the role of spontaneous emission. Specifically, section 2 is devoted to a brief review of the mechanism proposed in Ref. \cite{rey07}. Section 3 studies how inelastic reflection contributes to heating. Sections 4 and 5 address the question of the quantum limit to the cooling rate. Section 6 investigates whether cooling can survive if spontaneous emission is allowed in the driving source, yielding a positive answer. A corollary is that heat pumping remains possible if the ac source is replaced by a hot dynamic environment. Section 7 discusses some features of ac cooling and compares it to other proposed mechanisms. A summary is given in section 8.

\section{Nonadiabatic pumping of heat.}

The ac cooling mechanism proposed in Ref. \cite{rey07} is schematically depicted in Fig. \ref{eq:model-processes}.
An asymmetric resonant-tunneling structure is formed by two wells
each of which hosts two quasibound states. The four levels are
symmetrically disposed so that the energy difference is smaller in
the right (R) than in the left (L) well. On the other hand, the
difference between the two upper levels is taken to be the same as
that
between the two lower ones, both being equal to the driving frequency: $E_{2%
L }-E_{2R}=E_{1R}-E_{1L }=\hbar \Omega >0$. In those
conditions, electron transport is dominated by two processes: (i)
electrons in the R electrode with energy $E_{2R}$ are
inelastically transmitted
to the L electrode, where they enter with energy $E_{2L }=E_{2R%
}+\hbar \Omega $, and (ii) electrons in the left with energy $E_{1L
}$ are transmitted to the right while also absorbing a photon.
For simplicity we may
assume a common chemical potential $\mu =\mu _{L }=\mu
_{R}$. Then in the right lead one is
effectively replacing \textit{hot} electrons (with energy $\varepsilon >\mu $%
) by \textit{cold} electrons ($\varepsilon <\mu $), i.e. the right electrode is being \textit{cooled }at the
expense of heating the left electrode. This mechanism may be viewed as
the basis of a quantum refrigerator. Under
suitable conditions the two dominant transport mechanisms may cancel
each other yielding a vanishing electric current, which prevents
electrode charging.

It is common to refer to electrons as hot or cold depending on whether their energy is above or below the chemical potential. The entropy variation in an
infinitesimal process is given
by $TdS=dU-\mu dN$. For independent electrons, this translates into $%
TdS=(\varepsilon-\mu)dN$, where $\varepsilon$ is the energy of the
electrons being added ($dN>0$) or removed ($dN<0$). However, the temperature variation is rather given by $C_{V}dT=(\varepsilon -\sigma )dN$, where $C_{V}$ is the heat
capacity and $\sigma \equiv \mu -T\left( \partial \mu /\partial
T\right) _{n}$, with $n$ the particle density. Thus in elementary processes where $N$ varies, the changes in entropy and temperature are not proportional in general. However, they may be assumed to be proportional in the interesting case where, on average, ($\dot{N}=0$) \cite{rey07}.

The heat production rate in lead $\ell =L ,R$ \cite%
{siva86,mosk02,avro04,arra07}:
\begin{equation}
\dot{Q}_{\ell }=\sum_{q}(\varepsilon _{q}-\mu _{\ell })\dot{N}_{\ell
q}\;, \label{eq:Q-dot}
\end{equation}%
where $N_{\ell q}$ and $\varepsilon _{q}$ are the electron number and
energy of state $q$ in electrode $\ell $ of chemical potential $\mu
_{\ell }=\mu $.

We consider a quantum-well heterostructures where the electron potential in the
perpendicular $z$ direction has the piecewise constant form shown in
Fig.~\ref{eq:model-processes} while it is uniform in the parallel
$xy$ plane. In such a delocalized system, the independent-electron
approximation is generally adequate. The bottom of the right well is made to
oscillate as
\begin{equation}
V(t)=V_{0}+V_{\mathrm{ac}}\cos (\Omega t)\;, \label{eq:driving}
\end{equation}
while the
left well operates in phase opposition with the same amplitude and
frequency. We focus on transport through a single channel.

Electron transport properties can be described in terms of
scattering probabilities. Within a single-channel picture, the
electric current flowing
into lead R under ac driving is given by \cite{wagn99ab,kohl05,rey05}%
\begin{equation}
\dot{N}_{R}=\frac{1}{h}\sum_{k=-\infty }^{\infty }\int d\varepsilon %
\left[ T_{RL}^{(k)}(\varepsilon )f_{L }(\varepsilon )-T_{L
R}^{(k)}(\varepsilon )f_{R}(\varepsilon )\right] \,,
\label{eq:NR-dot}
\end{equation}%
where $f_{\ell }(\varepsilon )$ is the Fermi distribution in lead
$\ell $ and $T_{\ell \ell ^{\prime }}^{(k)}(\varepsilon )$ is the
probability for an electron to be transmitted from lead $\ell
^{\prime }$ to lead $\ell $ while its energy changes from
$\varepsilon $ to $\varepsilon +k\hbar \Omega $, $k$ being an
integer number. In this language, Eq.~\eqref{eq:Q-dot}
leads to \cite{rey07}
\begin{align}
\dot{Q}_{R}=\frac{1}{h}\sum_{k=-\infty }^{\infty }\int d\varepsilon %
\big[& (\mu _{R}-\varepsilon )T_{L R}^{(k)}(\varepsilon )f_{%
R}(\varepsilon )  \label{eq:QR-dot} \\
& +(\varepsilon +k\hbar \Omega -\mu _{R})T_{RL %
}^{(k)}(\varepsilon )f_{L }(\varepsilon )  \notag \\
& +k\hbar \Omega R_{RR}^{(k)}(\varepsilon )f_{R%
}(\varepsilon )\big]\;,  \notag
\end{align}%
where $R_{RR}^{(k)}(\varepsilon )$ is the probability
that an electron is reflected in lead R from energy $\varepsilon $
to $\varepsilon
+k\hbar \Omega $. For later use we note here that Eq. (\ref{eq:QR-dot}) can also be written as
\begin{eqnarray}
\dot{Q}_{R} &=&\frac{1}{h}\int \! \int d\varepsilon \! ~d\varepsilon ^{\prime }[-\varepsilon
T_{LR}(\varepsilon ^{\prime },\varepsilon )f_{R}(\varepsilon
)+\varepsilon ^{\prime }T_{RL}(\varepsilon ^{\prime },\varepsilon
)f_{L}(\varepsilon )  \nonumber \\
&&~~~~~~~~~~~~~~~~+(\varepsilon ^{\prime }-\varepsilon )R_{RR%
}(\varepsilon ^{\prime },\varepsilon )f_{R}(\varepsilon )]~,
\label{heat-1}
\end{eqnarray}%
where $\mu _{\rm R}\equiv 0$ and the scattering probabilities have been
rewritten $S_{ij}(\varepsilon ^{\prime },\varepsilon )\equiv \sum_{k=-\infty
}^{\infty }S_{ij}^{(k)}\delta (\varepsilon ^{\prime }-\varepsilon-\hbar k\Omega )$. As
formally both $\varepsilon $ and $\varepsilon ^{\prime }$ run over all real
values, we may assume the scattering probabilities $S_{ij}(\varepsilon
^{\prime },\varepsilon )$ to be zero where physically required.

\section{Heating due to inelastic reflection.}

The inelastic reflection term in Eq. (\ref{heat-1}) may be analyzed separately. For clarity we remove subindex $R$ in this section, since only the $R$ electrode is relevant:
\begin{eqnarray}
\dot{Q} &=&\frac{1}{h}\int \! \int d\varepsilon \! ~d\varepsilon ^{\prime }
(\varepsilon ^{\prime }-\varepsilon )
R(\varepsilon ^{\prime },\varepsilon )f(\varepsilon )~.
\label{inelastic}
\end{eqnarray}%
Next we prove that, in the presence of time-symmetric driving [$V(t)=V(-t)$
in Eq. (\ref{eq:driving})], $\dot{Q}>0$, i.e. inelastic reflection can only contribute to heating.

In the presence of time-reversal symmetry, one has (see e.g. Ref. \cite{kohl05})
\begin{equation}
R(\varepsilon ^{\prime },\varepsilon )=R(\varepsilon ,\varepsilon^{\prime } )~.
\label{TRsymmetry}
\end{equation}
Then Eq. (\ref{inelastic}) can be rewritten as
\begin{eqnarray}
\dot{Q} &=&\frac{1}{2h}\int \! \int d\varepsilon \! ~d\varepsilon ^{\prime }
(\varepsilon ^{\prime }-\varepsilon )R(\varepsilon ^{\prime },\varepsilon )
[f(\varepsilon )-f(\varepsilon^{\prime } )]~.
\label{inelastic-2-symm}
\end{eqnarray}
The decreasing monotonic character of the Fermi distribution $f(\varepsilon )$ guarantees
$(\varepsilon ^{\prime }-\varepsilon )[f(\varepsilon )-f(\varepsilon^{\prime } )]>0$ and therefore
\begin{equation}
\dot{Q}>0~.
\label{inel-posi}
\end{equation}
We conclude that inelastic reflection always contributes to heating.

The requirement of time-reversal symmetry seems to suggest that, in its absence [i.e. for $V(t)\neq V(-t)$], Eq. (\ref{inel-posi}) might be violated. This is unlikely to be possible, at least in a number of cases high enough to be important. It is true that a given signal $V(t)$ either increases or decreases the energy content of a closed systems. Therefore either $V(t)$ or $\bar{V}(t)\equiv V(-t)$ will decrease the system average energy.

Let us assume, for instance, that $\bar{V}(t)$ increases the energy while $V(t)$ decreases it. The signal $V(t)$ is guaranteed to ``cool'' only if it acts on exactly the density matrix that results from driving the system under the effect of $\bar{V}(t)$ after having started with a cool thermal distribution. In general, we may expect that, if $V(t)$ acts on a generic thermal state with similar energy content, the effect will be that of heating the system.

We conclude that the most general behavior is that inelastic reflection contributes to heating, although it can only be rigorously proved for time-symmetric driving. Thus any mechanism, such as that depicted in Fig.~\ref{eq:model-processes} must be efficient enough to overcome the general heating effect of electron reflection.

The main result of Ref. \cite{rey07} was that, despite the heating due to reflection, it is possible cool the R electrode within the scheme of Fig.~\ref{eq:model-processes}. Numerically exact results obtained with the transfer-matrix method \cite{wagn95} proved that $\dot{Q}_{R}$ can be negative even when left electrode is hotter than the right electrode \cite{rey07}. The study included the most interesting case where cooling takes place while the net electric current is zero.

\section{Quantum limit to surface cooling (electrons).}

One may wonder whether there is any fundamental
limit to the maximum cooling rate per quantum channel which would
play a role analogous to the quantum of
electric or thermal conductance ($e^{2}/h$ and $\pi ^{2}k_{B%
}^{2}T/3h$, respectively). It has been argued \cite{rey07} that the maximum cooling
rate should be achieved in an ideal setup where a metal at
temperature $T$ is connected through a totally transparent interface
to another metal at the same chemical potential but at zero
temperature. The result is the quantum of cooling power:
\begin{equation}
C_{Q}\equiv |\dot{Q}|_{\text{max}}=\frac{2}{h}\int_{0}^{\infty
}d\varepsilon
\,\varepsilon \,f(\varepsilon )=\frac{\pi ^{2}}{6}\frac{k_{B%
}^{2}T^{2}}{h}\;,  \label{eq:q-limit}
\end{equation}%
where $f(\varepsilon )\equiv \lbrack \exp (\varepsilon /k_{B%
}T)+1]^{-1}$ and $\pi ^{2}k_{B}^{2}/6h=473$ fW K$^{-2}$.
Following
information theory arguments, a similar result can be derived \cite%
{pend83,blen00}. Differentiation of (\ref{eq:q-limit}) yields the
quantum of thermal conductance. Below we prove that Eqs.
(\ref{eq:QR-dot}) and (\ref{heat-1}) satisfy (with $T_{R}=T$)%
\begin{equation}
\dot{Q}_{R}\geq -C_{Q}~,  \label{inequality}
\end{equation}%
for arbitrary electrodes (including $\mu _{L }\neq \mu
_{R}$) and arbitrary (i.e. not necessarily time-reversal
symmetric) driving, thus confirming
rigorously the intuitive idea that $C_{Q}$
is an upper bound to the cooling rate.

First we note that, exchanging
variables $\varepsilon $ and $\varepsilon ^{\prime }$ where necessary, Eq. (%
\ref{heat-1}) may be rewritten%
\begin{eqnarray}
h\dot{Q}_R &=&\int \! \int d\varepsilon \! ~d\varepsilon ^{\prime }~\varepsilon
~\{-[T_{LR}(\varepsilon ^{\prime },\varepsilon )+
R_{RR}(\varepsilon ^{\prime },\varepsilon )]f_R(\varepsilon )  \nonumber \\
&&~+T_{RL}(\varepsilon ,\varepsilon ^{\prime
})f_{L}(\varepsilon ^{\prime })+R_{RR}(\varepsilon ,\varepsilon
^{\prime })f_{R}(\varepsilon ^{\prime })\}~.  \label{heat-2}
\end{eqnarray}%
By unitarity, we have $\int d\varepsilon ^{\prime }~[T_{LR%
}(\varepsilon ^{\prime },\varepsilon )+R_{RR}(\varepsilon ^{\prime
},\varepsilon )]=1$, which yields a first term equal to
\begin{equation}
-\int d\varepsilon ~\varepsilon ~f_{ R}(\varepsilon )=-\int_{0}^{\infty
}d\varepsilon ~\varepsilon ~[2f_{ R}(\varepsilon )-1]~.  \label{heat-3}
\end{equation}%
For the second and third term in the integrand of (\ref{heat-2}) we use the
elementary identity $\int d\varepsilon ~\varepsilon ~F(\varepsilon
)=\int_{0}^{\infty }d\varepsilon ~\varepsilon ~[F(\varepsilon
)-F(-\varepsilon )]$ and rewrite its sum as%
\begin{eqnarray}
&&\int d\varepsilon ^{\prime }\int_{0}^{\infty }d\varepsilon ~\varepsilon
~[T_{RL}(\varepsilon ,\varepsilon ^{\prime })f_{ L}(\varepsilon
^{\prime })+R_{RR}(\varepsilon ,\varepsilon ^{\prime
})f_{ R}(\varepsilon ^{\prime })  \nonumber \\
&&~~~~~~~~~~~~~~-T_{RL}(-\varepsilon ,\varepsilon ^{\prime
})f_{ L}(\varepsilon ^{\prime })-R_{RR}(-\varepsilon ,\varepsilon
^{\prime })f_{ R}(\varepsilon ^{\prime })]  \nonumber \\
&\geq &-\int_{0~}^{\infty }d\varepsilon ~\varepsilon ~\int d\varepsilon
^{\prime }[T_{RL}(-\varepsilon ,\varepsilon ^{\prime })+R_{RR%
}(-\varepsilon ,\varepsilon ^{\prime })]~,  \label{heat-4}
\end{eqnarray}%
where, for the inequality, we have used $0\leq f\leq 1$. Invoking unitarity again, (\ref%
{heat-4}) becomes $-\int_{0~}^{\infty }d\varepsilon ~\varepsilon$, which
cancels the divergent term in (\ref{heat-3}). Finally, we obtain
\begin{equation}
\dot{Q}_{ R}\geq -\frac{2}{h}\int_{0}^{\infty }d\varepsilon ~\varepsilon
~f_{ R}(\varepsilon )~, \label{final-bound}
\end{equation}%
which proves our assertion.

We wish to emphasize that our proof of the inequality (\ref{inequality}) or (\ref{final-bound}) applies to an arbitrary driving setup. In particular, $\mu_{ L}$ can take any value and the driving signal does not have to be symmetric under time-reversal, i.e. it can be more general than the signal (\ref{eq:driving}).

\section{Quantum limit to bulk cooling (atoms).}

The quantum limit derived above may be intuitively understood as follows:
$k_{B}T$ is the maximum amount of heat that can be carried
away in an elementary
process. Such processes take place at a rate $\sim|\dot{Q}|/k_{B}T$%
, which cannot exceed $h/k_{B}T$ if one is to avoid
effective
heating caused by energy uncertainty. This results in $|\dot{Q}|\lesssim k_{%
B}^{2}T^{2}/h$, as given more precisely in Eqs.
\eqref{eq:q-limit}-\eqref{inequality}. This argument suggests that
$k_{B}^{2}T^{2}/h$ is also a quantum limit for the cooling
rate per active degree of freedom (one with characteristic energy scale
$\ll k_{B}T$) and as such could be relevant also to the bulk cooling of other particles such as e.g. atoms. In general we may write the internal energy of a quantum system as
\begin{equation}
U \simeq N_d k_B T \, ,
\label{UNdkT}
\end{equation}
where $N_d$ can be interpreted as the number of active degrees of freedom (those possessing an energy $\sim k_B T$). In a system where $U$ has an internal power-law dependence on $T$ ($U\sim T^{\alpha}$), one may identify $N_d\simeq C_V /\alpha k_B$.
In this language, the general form of the quantum limit would be
\begin{equation}
|\dot{E}|\lesssim k_{B}^{2}T^{2}/h\, ,
\label{U-dot}
\end{equation}
where $E=U/N_d\simeq k_BT$ is the energy per active degree of freedom. We may conclude
\begin{equation}
\dot{T}\gtrsim -k_B T^{2}/h\, ,
\label{T-dot}
\end{equation}
which is the central result of this section.

Let us now consider a gas of atoms moving under strong friction in an optical molasses. The magnitude of the stopping force experienced by an atom moving at speed $v$ is $\eta v$, where $\eta$ is the friction coefficient. Thus the rate at which the kinetic energy (per component) decreases is
\begin{equation}
\dot{E}=-\eta v^2 \approx - \eta k_B T / m\, ,
\label{E-dot}
\end{equation}
where $m$ is the atom mass. The combination of (\ref{T-dot}) and (\ref{E-dot}) leads to the inequality
\begin{equation}
T \gtrsim T_{\rm min} = \frac{h}{k_B}\frac{\eta}{m}
\label{T-min}
\end{equation}
Thus the mere existence of a quantum limit to the cooling rate, as expressed in (\ref{U-dot}) or (\ref{E-dot}), already implies the existence of a minimum achievable temperature.

For atom laser cooling, other arguments lead to identify the recoil temperature $T_r=\hbar^2 k_L^2/mk_B$ as the minimum temperature, where $k_L$ is the laser light wave number \cite{metc99}. Since on the other hand, $\hbar k_L^2 /4$ is shown to be the highest possible value of $\eta$ \cite{metc99}, we conclude that the inequality (\ref{T-min}) is essentially guaranteed to be satisfied in laser cooling setups.

\section{Dynamic environment. Spontaneous emission.}

So far we have assumed a semiclassical driving (\ref{eq:driving}) by a
source without internal degrees of freedom. We may wonder whether heat
pumping remains possible if the driving field is allowed to have some
internal dynamics whose main signature would be the possibility of
spontaneous emission. The coherent driving would appear as the classical
limit of the oscillator (here, the photon mode) prepared in a coherent state of
large amplitude.

First we should note that the posed problem has no exact solution. This limitation has
for long precluded an exact numerical study of the effect of phonons
(or photons in the present case) on the transport of electrons in
nanostructures in the presence of arbitrary one-electron scattering. The essential difficulty appears when one attempts to include
simultaneously (i) inelastic scattering due to a dynamic environment (as
opposed to a semiclassical ac source), (ii) electron scattering in an
arbitrary nanostructure (in particular, beyond the tunnelling limit), and (iii) the
Pauli exclusion principle (Fermi statistics). It was already argued in Ref.  \cite{sols92}
that an exact combination of ingredients (i), (ii), and (iii) above is not
possible [even (i) and (ii), akin to the polaron problem, has no exact solution]. It was noted, however, that the situation becomes simpler in a number of
limiting cases. One of them is that where the set of initial states is
identical to the set of final states and the coupling to the environment is treated perturbatively. A typical example is given by the tunnelling
limit, where stationary waves span the set of both incoming and outgoing
scattering channels. Here we focus in this limit because it permits an
analytical study.

\begin{figure}[tbp]
\begin{center}
\includegraphics[angle=0,width=.40\textwidth]{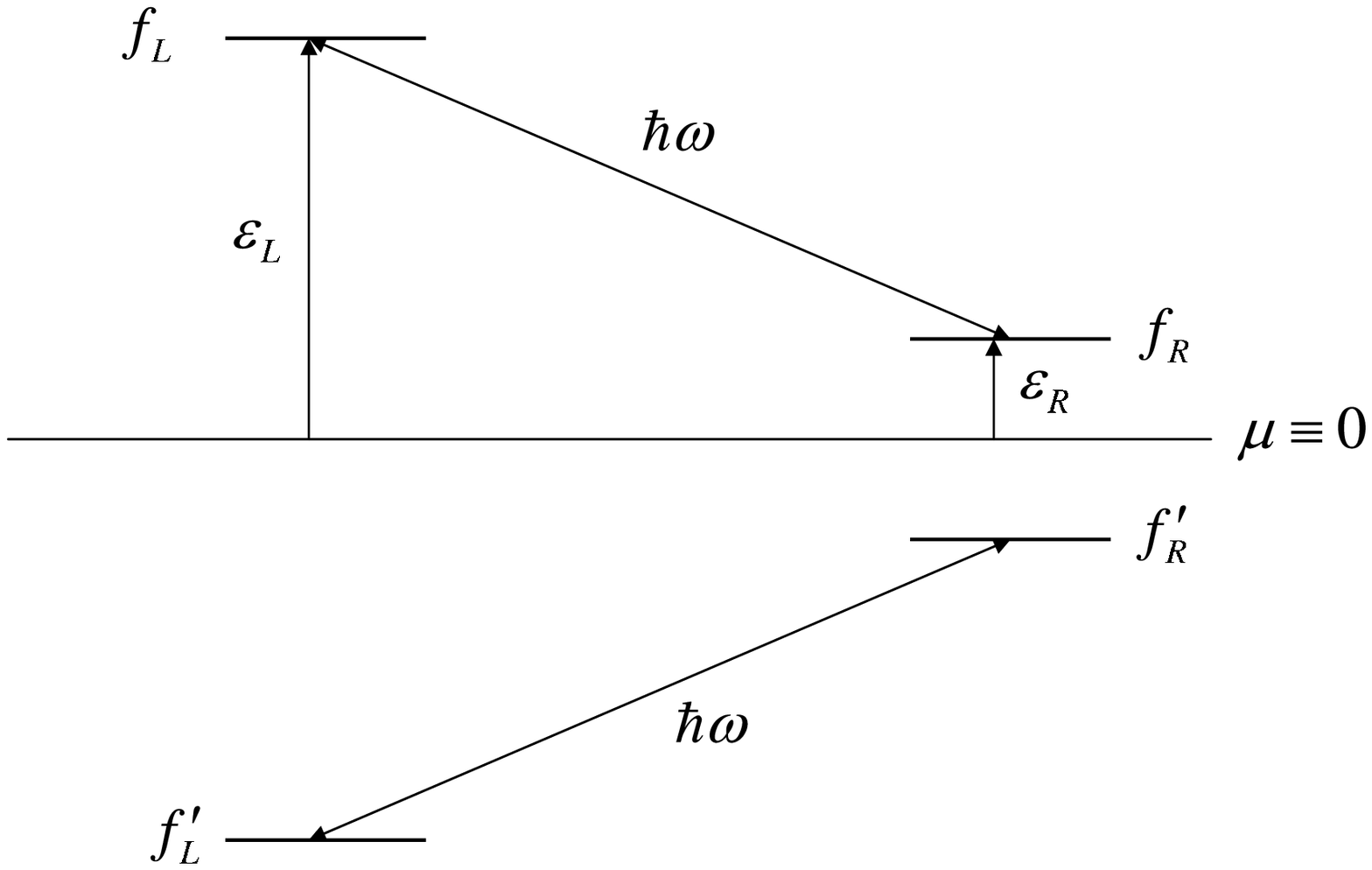}
\end{center}
\caption{Schematic diagram of energy levels.}
\label{schematic}
\end{figure}

To simplify the discussion further we assume that the setup of Fig. \ref%
{eq:model-processes} imposes a strict filter on the electron energies, so that only
the precise energies indicated in Fig. \ref{schematic} can contribute to
transport. The relation between the energies in Figs. \ref{eq:model-processes} and
\ref{schematic} is straightforward. In the (assisted) tunnelling limit and in the presence of an oscillator field of frequency $\omega$, two different levels which differ in energy by $\hbar \omega$ are connected in such a way that, if an electron starts in one level, on can calculate the probability per unit time that the electron jumps to the other level. A Fermi golden rule calculation would involve an effective tunnelling matrix element, Fermi-Dirac occupation factors, and a delta function ensuring conservation of energy (which, to be well defined, requires that the oscillator field or the electrons have a continuous density of states). For convenience, we neglect detailed prefactors and capture the essence of the total electric current through the expression
\begin{eqnarray*}
I &=&I_{1}+I_{2} \\
I_{1} &=&-nf_{R}(1-f_{L})+(n+1)f_{L}(1-f_{R}) \\
I_{2} &=&-(n+1)f_{R}^{\prime }(1-f_{L}^{\prime })+nf_{L}^{\prime
}(1-f_{R}^{\prime })~.
\end{eqnarray*}%
We adopt the convention that $I>0$ if current flows from
left to right, so that it is proportional to $\dot{N}_{R}$ in Eq.
(\ref{eq:NR-dot}). Here $n$ is the (large)
number of quanta in the field mode yielding the ac driving.$\ \ I_{1}$ is
the current through the upper channel of Fig. \ref{schematic}, and $I_{2}$
that through the lower channel. Primes indicate that the distributions are
evaluated at the lower energies. For instance, $f_{R}$ stands for $%
f_{R}(\varepsilon _{R})$ while $f_{R}^{\prime }$ represents $%
f_{R}(-\varepsilon _{R})$ (we assume that the level structure in
Fig. \ref{schematic} is symmetric).

Using the identities
\begin{equation}
f_{L}^{\prime }=1-f_{L}~~,~\ \ ~~\ ~\ ~~f_{R}^{\prime }=1-f_{R}~~,
\label{symmetric}
\end{equation}%
$I_{2}$ becomes%
\[
I_{2}=-(n+1)(1-f_{R})f_{L}+n(1-f_{L})f_{R}=-I_{1}
\]%
so that $I=0$, as expected for the symmetric case. A first conclusion is
that zero electric current is possible in the presence of spontaneous
emission. This should be possible in the general case (i.e. for generic
temperatures, chemical potentials and level structure), as suggested by the
following argument: Starting from $I=0$ in the symmetric case, one may depart from
the symmetric limit by changing some parameters while compensating that
change with other parameters so that $I$ remains zero.

We focus on \textit{heat transport} due to electron transmission, i.e. we
neglect processes where the electron stays in the same electrode (inelastic
reflection). It was stated in Ref. \cite{rey07}, and has been proved in
section 3, that inelastic reflection under coherent time-symmetric driving contributes only to heating.

Like in Ref. \cite{rey07}, we focus on the heat production at $R$, assume $%
\mu _{R}=0$, and take the two resonant levels in the right well
symmetrically disposed around the energy origin, at a distance $\varepsilon
_{R}=\varepsilon >0$. Then the heat production is%
\begin{eqnarray*}
\dot{Q}_{R} &=&\dot{Q}_{1}+\dot{Q}_{2} \\
\dot{Q}_{1}/\varepsilon  &=&-nf_{R}(1-f_{L})+(n+1)f_{L}(1-f_{R}) \\
\dot{Q}_{2}/\varepsilon  &=&(n+1)f_{R}^{\prime }(1-f_{L}^{\prime
})-nf_{L}^{\prime }(1-f_{R}^{\prime })
\end{eqnarray*}%
As before, we employ the convention that $\dot{Q}_{R}>0$ if heat is given to
electrode $R$ and $<0$ if heat is extracted from $R$.\ \ We assume the
symmetric case again [Eq. (\ref{symmetric})] and obtain $\dot{Q}_{1}=\dot{Q}%
_{2}$, so that%
\begin{equation}
\dot{Q}_{R}/2\varepsilon  =-n(f_{R}-f_{L})+f_{L}(1-f_{R}) \, ,  \label{general}
\end{equation}
which is the central result of this section.
Next we study some particular limits.

(1) \textit{Classical source.} It corresponds to $n \gg 1$, so that
$\dot{Q}_{R}\propto n(f_{L}-f_{R})$, where the omitted prefactor becomes small to yield a finite cooling rate. Cooling is guaranteed provided
$f_{L}<f_{R}$. If $L$ is hotter than $R$, refrigeration of $R$ is possible provided $\varepsilon_L$ is placed sufficiently high (and the frequency is adpated correspondingly to preserve the resonant condition $\varepsilon_L=\varepsilon_R+\hbar\omega$). This classical limit has been implicitly assumed in Ref. \cite{rey07} and in sections 2-4 of the present paper.

(2) \textit{Identical wells (zero frequency)}. If $\varepsilon _{L}\rightarrow\varepsilon _{R}$%
, then $\omega \rightarrow 0$ and $n \rightarrow \infty$. The photons become increasingly soft. For coherent driving, this is a delicate limit that has been studied in e.g. Ref. \cite{hone97}. Here we are only interested in the fact that, as $\omega$ vanishes, the effective $n$ becomes large.
As a result, $\dot{Q}_{R}\propto n(f_{L}-f_{R})$, i.e. we
obtain the simple result that $\dot{Q}_{R}>0$ if $R$ is colder than $L$, and
$<0$ in the opposite case: in the effective absence of driving, heat spontaneously flows from hot to cold.

(3) \textit{Cold source.} Then $n=0$. Thus $\dot{Q}_{R}\propto f_{L}(1-f_{R})>0
$. When taking $n=0$ we are implicitly assuming that the dynamic environment
is there but at zero temperature. Heating would occur due to spontaneous
emission across the interface, to the extent that it is possible [i.e. if
both factors, $f_{L}$ and $(1-f_{R})$, are nonzero]. In this case inelastic
reflection, which we do not consider explicitly in this section, would
produce cooling, since the $R$ electrode would be cooled by the dynamical
coupling to a zero-temperature source or environment ($T_{S}=0$).

Returning to the general case, we conclude from (\ref{general}) that the $R$
electrode is cooled ($\dot{Q}_{R}<0$) when%
\[
n(f_{R}-f_{L})>f_{L}(1-f_{R})
\]
If $f_{L}>f_{R}$ (i.e. if $L$ is too hot or $\varepsilon _{L}$ is too low)
this is not possible. We note that these distributions can be related to the
electrode temperatures,%
\[
f_{L}=\frac{1}{e^{\beta _{L}\varepsilon _{L}}+1}~~~,~~f_{R}=\frac{1}{%
e^{\beta _{R}\varepsilon _{R}}+1}~~.~
\]%
So, by placing $\varepsilon _{L}$ sufficiently high up we can have $%
f_{L}<f_{R}$ and yet $T_{L}>T_{R}$ (see Fig. \ref{schematic}). This
\textquotedblleft non-trivial cooling\textquotedblright , whereby heat is
extracted from colder $R$, is the most interesting one. Hereafter we focus on this case ($f_L < f_R$).

Whether the source is thermal or coherent (semiclassical) we can always
define and effective source temperature such that%
\[
n=\frac{1}{e^{\beta _{S}\hbar \omega }-1}~.
\]%
We can also define an effective occupation number
\[
\bar{n}\equiv \frac{f_{L}(1-f_{R})}{f_{R}-f_{L}}>0~,
\]
so that the general result (\ref{general}) can be rewritten as
\[
\dot{Q}_{R}/2\varepsilon = - (n-\bar{n})(f_{R}-f_{L})~.
\]
An effective temperature $\bar{T}$ can also be defined such
that%
\[
\bar{n}=\frac{1}{e^{\bar{\beta}\hbar \omega }-1}=\frac{1}{e^{\beta
_{L}\varepsilon _{L}-\beta _{R}\varepsilon _{R}}-1}~.
\]%
For the second equality we have used $\varepsilon _{L}-\varepsilon
_{R}=\hbar \omega $, so that
\[
\bar{\beta}=\frac{\beta _{L}\varepsilon _{L}-\beta _{R}\varepsilon _{R}}{%
\varepsilon _{L}-\varepsilon _{R}}
\]%
Then the cooling condition can be expressed as%
\begin{eqnarray}
\dot{Q}_{R} &<&0~~~\text{for \ \ }T_{S}>\bar{T}~~,~\ \ n>\bar{n} \, .
\label{criterion}
\end{eqnarray}%
We conclude that there is cooling if the amplitude or temperature of the external source is sufficiently high.

We end by noting that, given a semiclassical ac source of amplitude $V_{%
\text{ac}}$, there is not a \ unique way to determine the effective $n$ for
the source, and in particular the $1/n\ll 1$ correction stemming from spontaneous emission. In
order to know $n$, one should have complete information on the
electromagnetic signal which is providing that ac driving of amplitude $V_{%
\text{ac}}$ (including its properties outside the sample region).
Fortunately, that procedure is not necessary in practice, since one deals
directly with the ac source as a semiclassical time-dependent perturbation
without having to invoke the effective value of $n$.

\section{Discussion.}

The study of the previous section proves that a coherent, semiclassical
source is not essential to pump heat from cold $R$ to hot $L$. A
sufficiently hot thermal source can provide the same effect. This is consistent with the
results of Ref. \cite{peko07}. For the purpose of pumping of electric
current (which would be nonzero for a non-symmetric level structure), this
is also consistent with the concepts of photovoltaic conversion where the
sun plays the role of the signal, as discussed e.g. in Ref. \cite{reim02}.

In the particular case $T_{L}\simeq T_{R}\equiv T$ \ we have $\bar{T}\simeq T$. Then (\ref{criterion}) allows us to state that a necessary condition for a thermal bath to induce heat pumping is that its temperature is higher than that of the electron
system to be cooled.

The analogy between the present analysis and the results of Ref. \cite%
{peko07} is worth discussing further. Both have in common the presence of a
dynamic environment acting as an effective driving source. Interestingly,
they also have an essentially similar electronic level structure. The low
value of $\varepsilon _{R}$ plays the role of the continuum of low-energy
excitations in the normal metal. The high value of $\varepsilon _{L}$ plays
a role similar to that of the superconducting gap $\Delta $. The symmetry
around the chemical potential $\mu =0$ is equivalent to the electron-hole
symmetry at a NS interface. Therefore the present model captures in
a simple way the essence of pump heating at an asymmetric interface: it
suffices to have a sufficiently hot source acting on the interface and a
good gap in the hot electrode (and sufficiently small heating due to
inelastic reflection). This picture permits a qualitative understanding of
the Brownian refrigerator discussed in \cite{peko07}.

The similarities between the mechanism of Ref. \cite{peko07} and those of Ref. \cite{rey07} and the present paper permit to identify the presence of an effective gap in the hot electrode as a useful element in the design of a heat pumping setup.

In Ref. \cite{rey07} the question was discussed of whether, in a setup like that of Fig.~\ref%
{eq:model-processes}, it is possible to approach the quantum limit to the cooling rate. It was concluded that cooling rate is maximized when $\varepsilon_R$ of Fig. \ref{schematic} (or $E_{2R}-\mu$ of Fig. \ref{eq:model-processes}), the resonance width, and the temperature, are all comparable. In such a case the cooling rate is only limited by the height of the transmission peak at the resonance, i.e. by the value of the maximum electron transmission probability. On the other hand, the linewidth of the resonance poses a limit to the energy resolution and thus a lower limit to the minimum achievable temperature. Although the work of Ref. \cite{peko07}, complemented by the discussion in Section 6, seems to suggest that a resonant structure may not be essential to produce cooling (in the sense that a gap in the hot electrode may produce the same effect), it is hard to figure out how one could get close to the quantum limit without resorting to resonances that would permit a maximum transmission close to unity.

The quantum refrigerator which we has been investigated here and in Ref. \cite{rey07}
may be viewed as a realization of Maxwell's demon \cite{maxw1871,parr02} as it selectively
lets hot electrons out and cold electrons in. The
required energy is provided by the external ac (or thermal) source which, combined
with the spatial asymmetry of the structure, rectifies electron
motion.

It was also noted in Ref. \cite{rey07} that a non-resonant mechanical mismatch at the
interface could be introduced to prevent phonons from short-circuiting electron
cooling, assuming that electron-phonon coupling is strong enough to pose a serious threat to electron cooling.

\section{Conclusions.}

We have discussed a mechanism for electron cooling based
on the coherent control of electron ac transport and which can
operate at zero average electric current. The feasibility of such an electron heat pump
was numerically demonstrated in Ref. \cite{rey07}. Motivated by the understanding of this refrigeration concept, we have explored some general questions of quantum cooling. First we have proved that, in the presence of time-reversal symmetry, inelastic reflection can only contribute to heating. Invoking arguments of electron quantum transport theory, we have derived a rigorous quantum limit to the cooling rate of electrons in the presence of arbitrary (i.e. not necessarily time-symmetric) driving. On the basis of qualitative arguments, we have derived a similar upper bound to the bulk cooling rate of atoms and have shown that it is guaranteed to be satisfied in laser cooling setups. Within the tunneling approximation, we have also discussed the possible replacement of the coherent ac source by an external thermal bath and found that the cooling effect is preserved if the bath temperature is high enough. We have identified the existence of minimum electron and hole energies to enter the hot electrode as a generic useful feature for the refrigeration of the cold electrode, and have identified it as a concept that could be exported to a variety of cooling setups.

\section*{Acknowledgements}
$\\$I would like to thank P. H\"anggi, S. Kohler, M. Rey and M. Strass
for helpful discussions and for the collaboration that led to Ref. \cite{rey07}.
This research has been supported by MEC (Spain) Grant No. FIS2007-65723.


%

\end{document}